\documentclass[onecolumn,a4paper]{revtex4}
\usepackage[mathscr]{euscript}
\usepackage{ulem}
\usepackage{caption}
\usepackage{multirow}
\usepackage{booktabs,tabularx}
\usepackage{threeparttable}
\usepackage{floatrow}
\usepackage[font=small,labelfont=bf,tableposition=top]{caption}
\usepackage{booktabs}
\usepackage{threeparttable}
\usepackage{graphicx}
\usepackage{subfigure}
\usepackage{epstopdf}
\usepackage[colorlinks=true,linkcolor=red]{hyperref}
\usepackage{geometry}
\usepackage{setspace}

\newcommand{\bea}{\begin{eqnarray}}
\newcommand{\eea}{\end{eqnarray}}
\newcommand{\beq}{\begin{equation}}
\newcommand{\eeq}{\end{equation}}

\def\/{\over}

\begin{document}

\begin{spacing}{1.4}

\title{Probing long-range properties of vacuum altered by \\uniformly accelerating two spatially separated Unruh-DeWitt detectors}

\author{Shijing Cheng$^{1,2,3}$, Wenting Zhou$^{4,}$\footnote{Corresponding author: zhouwenting@nbu.edu.cn} and Hongwei Yu$^{1,}$\footnote{Corresponding author: hwyu@hunnu.edu.cn}}

\affiliation{
$^{1}$
Department of Physics and Synergetic Innovation Center for Quantum Effect and Applications, 
Hunan Normal University, Changsha, Hunan 410081, China\\
$^{2}$School of Fundamental Physics and Mathematical Sciences, Hangzhou Institute for Advanced Study, UCAS, Hangzhou 310024, China\\
$^{3}$School of Physical Sciences, University of Chinese Academy of Sciences, No.19A Yuquan Road, Beijing 100049, China\\
$^{4}$Department of Physics, School of Physical Science and Technology, Ningbo University, Ningbo, Zhejiang 315211, China
}

\begin{abstract}
In a quantum sense, vacuum is not an empty void but full of virtual particles (fields). It may have long-range properties, be altered, and even undergo phase transitions. It is suggested that long-range properties of a quantum vacuum may be probed by distributing matter over a large spatial volume. Here, we study a simplest example of such, i.e., two uniformly accelerated Unruh-DeWitt detectors which are spatially separated, and examine the inter-detector interaction energy arising from the coupling between the detectors and fluctuating fields to see if novel phenomena related to the long-range properties emerge of a vacuum altered by uniformly accelerating two spatially separated detectors through it. Our results show that when the inter-detector separation is much larger than the thermal wavelength of the Unruh thermal bath, the inter-detector interaction displays a completely new behavior, which, as compared with that of the inertial detectors, is surprisingly exclusively acceleration-dependent, signaling a new phase of the vacuum in which its imprint as seen by two inertial observers seems to be completely wiped out. Moreover, we demonstrate that the inter-detector interaction in the near region can be significantly enhanced by the accelerated motion in certain circumstances, and with  two Rydberg atoms as the detectors,
the acceleration required for an experimentally detectable enhancement of the interaction energy can be $10^5$ times smaller than that required for the detection of the Unruh effect.
\end{abstract}

\maketitle

\section{Introduction}

It is now well recognized that, in sharp contrast to a classical vacuum, a quantum vacuum is not an empty void but full of transient virtual evanescent particles and antiparticles as a result of vacuum fluctuations that inevitably exist as necessitated by the Heisenberg uncertainty principle. The vacuum fluctuations are ubiquitous in nature and responsible for various phenomena, such as the Casimir effect~\cite{Casimir48}, the Lamb shift~\cite{Lamb47,Scully10}, the dispersion force between atoms~\cite{Casimir-Polder48,Craig98,Salam10} and the Brownian motion of a charged particle in vacuum~\cite{Gour99,Yu04}.

The connotation of quantum vacuum has been remarkably enriched by extensive studies on the interaction between matter and fluctuating quantum vacuum fields. 
The Unruh effect, which 
{attests that a uniformly accelerated observer perceives the  vacuum of an inertial observer as a thermal bath at a temperature proportional to its proper acceleration $a$, $T_U={a\/2\pi}$,} is one of the striking phenomena which results from matter-vacuum interaction and has profoundly  changed our understanding of quantum vacuum. 
{This effect was originally discovered by Unruh~\cite{Unruh76} by examining the response rate of a uniformly accelerated two-level point detector, called the Unruh-DeWitt (UDW) detector~\cite{DeWitt}, in the monopole interaction with a massless scalar field in vacuum.
Later, it has also been derived via the transition rates~\cite{Birrell82,Audretsch94,Bilge98,Yu05,Barbado20,Zhu06,Zhou12} and the radiative energy shifts~\cite{Audretsch95,Rizzuto07,Rizzuto09,Passante98,Zhu10} of the accelerated detector.
With a  detector alternatively modeled by a harmonic oscillator or a charged particle, a series of researches have been devoted to the issue of the Unruh radiation which has realistic significance in light of experimental proposals on the detection of the Unruh effect.
It is found that the quantum radiation is in general nonvanishing~\cite{Lin17,Raval97,Iso11,Oshita15,Lin06} [just to name a few]; however, in the special case of $(1+1)$-dimensional Minkowski spacetime, a uniformly accelerated detector produces no quantum radiation under equilibrium conditions as a result of quantum interference~\cite{Raine91,Raval96,Ford06}.
Moreover, there have also been studies on the Berry phase~\cite{Martin11,Hu12} and the thermalization~\cite{Brenna13} of  uniformly accelerated detectors. }

{The aforementioned studies on the Unruh effect itself and the related phenomena}
{all indicate} that the quantum vacuum as viewed by an accelerated observer is altered by the accelerated motion as compared with that viewed by an inertial observer, 
{and  novel properties of the acceleration-altered quantum vacuum are naturally manifested.}
Let us also note that another well-known example of alteration of a quantum vacuum is the Schwinger pair production as a result of an external strong electric field~\cite{Schwinger51}. The study on possible means to alter the properties of the vacuum not only is of paramount theoretical significance but may also be of vital importance to possible new technologies dubbed as vacuum engineering, a term which was evidently first coined by Lee, who anticipated 
{that vacuum may have both local and long-range coherent properties} and ``{\it if indeed we are able to alter the vacuum, then we may encounter some new phenomena, totally unexpected}~\cite{Lee}".
{It is expected that a single detector only explores local properties of an altered quantum vacuum. For  long-range ones, one may resort to a system of detectors which are distributed over a large spatial volume as suggested by Lee~\cite{Lee}, and one simplest such system should be that of two spatially separated detectors.}

In this paper, we plan to explore 
{long-range coherent properties of an altered quantum vacuum as a result of uniformly accelerating two spatially separated detectors through it. 
} 
{For this purpose, we concretely calculate the interaction energy between such two ground-state UDW detectors to see how the alteration of the vacuum caused by the uniform acceleration affects the inter-detector interaction.  We aim to search if and in what circumstances one can dramatically alter the properties of the vacuum and novel phenomena related to the long-range properties of the altered vacuum can emerge.
Here, we also note that other physical traits of the two-detector system could also be good candidates  to explore long-range properties of an acceleration-altered quantum vacuum.
There have been extensive studies showing that the accelerated motion can significantly affect the behaviors of the interatomic resonance interaction~\cite{Rizzuto16,Zhou161} and the collective transition rates~\cite{Zhou20,Zhou201} of two entangled atoms, as well as the dynamics of quantum entanglement between two detectors~\cite{Reznik03,Benatti04,Lin08,Lin10,Hu15,Ostapchuk12,Zhang07,Cheng18,ZhouY21}.
The effect on the entanglement dynamics between two coupled detectors  by the direct coupling of detectors~\cite{Hsiang15} as well as the  vacuum fluctuation induced coupling~\cite{Chen22} has  been further studied}.

\section{Effects of acceleration-induced-alteration in quantum vacuum on the interaction between two UDW detectors}

\begin{figure}
\centering
\includegraphics[width=6cm]{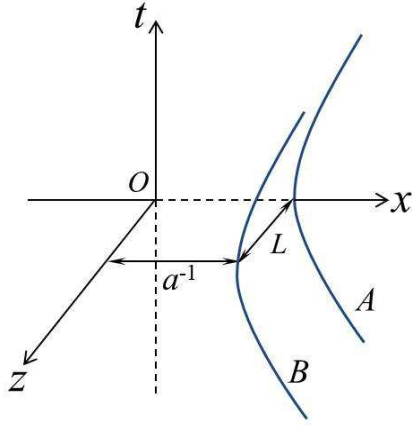}
\caption{\scriptsize 
{Spacetime diagram of two detectors synchronously and uniformly accelerated with a constant interatomic separation $L$ perpendicular to the constant proper acceleration $a$. The blue lines represent the trajectories of the two detectors respectively.}}
\label{acc}
\end{figure}

We consider that two UDW detectors labeled respectively by $A$ and $B$ with a constant separation $L$ are synchronously and uniformly accelerated with an acceleration $a$ in vacuum in a direction perpendicular to the separation, 
{and thus their proper times can be denoted by a same $\tau$ ({see Fig. \ref{acc} for the spacetime diagram}). Then} their trajectories, in terms of the proper time $\tau$, are depicted by
\bea
t_{A,B}(\tau)=\frac{1}{a}\sinh{(a\tau)},\ \ x_{A,B}(\tau)=\frac{1}{a}\cosh{(a\tau)},\ \ y_{A,B}(\tau)=0,\ \ z_A(\tau)=0,\ z_B(\tau)=L\;.
\label{trajectories}
\eea

The detectors are assumed to be of  two levels with energy gap $\omega_{\xi}$ $(\xi=A,B)$, and the lower and upper energy states  are denoted by $|g_{\xi}\rangle$ and $|e_{\xi}\rangle$, respectively.
We  consider that the 
{detectors are in their ground states} and are weakly coupled to the fluctuating massless scalar fields in vacuum, and the Hamiltonian of the ``detectors+field'' system is given by
\bea
H(\tau)&=&\omega_A R^A_{3}(\tau)+\omega_B R^B_{3}(\tau)+\int d^3{\mathbf{k}}\ \omega_{\mathbf{k}}a^{\dag}_{\mathbf{k}}(t)a_{\mathbf{k}}(t)\frac{dt}{d\tau}\nonumber\\
&&+\mu R^A_{2}(\tau)\phi(x_A(\tau))+\mu R^B_{2}(\tau)\phi(x_B(\tau))\;,
\label{total}
\eea
where {$R^{\xi}_{3}=\frac{1}{2}(|e_{\xi}\rangle\langle e_{\xi}|-|g_{\xi}\rangle\langle g_{\xi}|)$ and $R^{\xi}_{2}=\frac{i}{2}(|g_{\xi}\rangle\langle e_{\xi}|-|e_{\xi}\rangle\langle g_{\xi}|)$},
$\mathbf{k}$ stands for the wave vector of the scalar field modes, $a^{\dag}_{\mathbf{k}}(t)$ and $a_{\mathbf{k}}(t)$ are respectively the creation and annihilation operators with momentum $\mathbf{k}$ of the scalar field $\phi(x)$, and $\mu$ is a small coupling constant. Unless stated otherwise, we adopt, throughout the paper, the natural units with $\hbar=c=k_B=1$. 

{As a result of the interaction between each detector and the fluctuating field in vacuum, a monopole is induced in the detector and meanwhile a radiative field is emitted, which then acts on the other detector and thus an inter-detector interaction is generated.
Here w}e use the 
formalism proposed by Dalibard, Dupont-Roc and Cohen-Tannoudji 
{[DDC formalism]}~\cite{DDC82,DDC84} to calculate the 
interaction 
{between two ground-state detectors}. 
{The DDC formalism is a powerful approach to studying the evolution of a small system in  interaction between a large reservoir, which provides a clear separation for the contributions of reservoir fluctuations and the radiation reaction of the small system. During the past decades, it has been widely utilized for investigations on matter-field interaction, such as the spontaneous excitation and energy shifts of atoms in acceleration~\cite{Audretsch94,Zhu06,Zhou12,Passante98,Rizzuto09,Zhu10,Yu05,Audretsch95,Rizzuto07} and  static atoms in a thermal bath~\cite{Tomazelli03}, and the effects of the nontrivial topology of the cosmic string spacetime as well as the Gibbons-Hawking effect of the de Sitter spacetime on the radiative properties of atoms~\cite{Zhu07,Zhou10,Zhou16}, which are second order perturbation effects. Since we are discussing the interaction between two ground-state detectors which is of the fourth order, we should first generalize the DDC formalism from its original second order to the fourth order}. 
Although such a generalization was attempted~\cite{Noto14,Menezes17,Noto16}, a complete and correct fourth-order DDC formalism has been given by us in Ref.~\cite{Zhou21} with a detailed derivation only recently. 

Following the fourth-order DDC formalism (see Appendix.~\ref{derivation}), we find that the contribution of the vacuum fluctuations (vf) to the interaction energy of the detectors moving along trajectories, Eq.~(\ref{trajectories}), is given by
\bea
(\delta E)_{vf}=-\frac{\mu^4}{64\pi^4}\int_0^{\infty}d\omega_1\int_0^{\infty}d\omega_2\biggl(1+{2\/{e^{\omega_1/T_U}-1}}\biggr)
\frac{\omega_A\omega_B\omega_2 f(\omega_1,a,L) f(\omega_2,a,L)}{(\omega_2^2-\omega_1^2)(\omega_1^2-\omega_A^2)(\omega_1^2-\omega_B^2)}\;,
\label{vf}
\eea
where $f(\omega_1,a,L)\equiv{\sin\left({2\omega_1\/a}\sinh^{-1}({aL\/2})\right)\/L\sqrt{1+{1\/4}a^2L^2}}$, 
and that  of the radiation backreaction (rr) by
\bea
(\delta E)_{rr}&=&
-\frac{\mu^4}{64\pi^4}\int_0^{\infty}d\omega_1\int_0^{\infty}d\omega_2\frac{(\omega_1+\omega_A+\omega_B)f(\omega_1,a,L)f(\omega_2,a,L)}{(\omega_1+\omega_A)(\omega_1+\omega_B)(\omega_A+\omega_B)}
\biggl({1\/{\omega_1+\omega_2}}-{1\/{\omega_1-\omega_2}}\biggr)\nonumber\\&&
-\frac{\mu^4}{16\pi^4}\int_0^{\infty}d\omega_1\int_0^{\infty}d\omega_2
\frac{\omega_A\omega_B\omega_2 f(\omega_1,a,L) f(\omega_2,a,L) }{(\omega_2^2-\omega_1^2)(\omega_1^2-\omega_A^2)(\omega_1^2-\omega_B^2)(e^{\omega_1/T_U}-1)}-(\delta E)_{vf}\;.
\label{rr}
\eea
The sum of them then gives rise to the interaction energy, $\delta E=(\delta E)_{vf}+(\delta E)_{rr}$.
{In obtaining the above results, we have assumed the interaction time interval $\Delta\tau$ to be much larger than the correlation time of the fields in vacuum adapted to the system of two two-level detectors and much less than the relaxation time of the detectors for the DDC approach to be applicable~\cite{Cohen86}, and for computational convenience, we have treated $\Delta\tau=\tau-\tau_0$ in Eqs.~(\ref{2vf}) and (\ref{2rr}) as infinity.}
To explore the effects of the acceleration-induced alteration of the quantum vacuum on the inter-detector interaction energy, we, in the following,  first discuss  the inter-detector interaction energy in the inertial case and then in the acceleration case. For brevity, we assume $\omega_A=\omega_B=\omega$.

\subsection{Inertial case}

For two inertial detectors ($a=0$), 
{analytical results of} the vf- and rr-contribution 
{to the interaction energy of two detectors with an arbitrary separation $L$ are difficult to derive; however,} in the near or far region, where 
{$L$ is much shorter or larger than the transition wavelength of the detectors, $\lambda=2\pi\omega^{-1}$, approximate{ly} analytical results are obtainable}.

\subsubsection{
{Inter-detector interaction energy in the near region}}

In the near region where $L\ll\lambda$, the vf- and rr-contributions to the inter-detector interaction energy are approximated by
\bea
(\delta E)_{vf}\simeq\frac{\mu^4}{256\pi^3L}\;,\quad\quad\quad\label{inert-vf}\\
(\delta E)_{rr}\simeq-\frac{\mu^4}{512\pi^2\omega L^2}+\frac{\mu^4}{256\pi^3L}\;,\label{inert-rr}
\eea
which respectively result in a repulsive force proportional to $L^{-2}$ and an attractive force proportional to $L^{-3}$ between the two detectors. Obviously, $|\delta E|_{rr}\gg|\delta E|_{vf}$, 
and thus the rr-contribution dominates over the vf-contribution. As a result,
\beq
\delta E\simeq(\delta E)_{rr}\simeq-\frac{\mu^4}{512\pi^2\omega L^2}+\frac{\mu^4}{128\pi^3L}\;,\label{iner-near}
\eeq
and the two detectors attract each other with a force proportional to $L^{-3}$.

\subsubsection{
{Inter-detector interaction energy in the far region}}

In the far zone where $L\gg\lambda$, the vf- and rr-contributions are approximated by
\bea
(\delta E)_{vf}\simeq\frac{\mu^4\sin{(2\omega L+\theta)}}{512\pi^2L}-\frac{\mu^4}{512\pi^3\omega^2 L^3}\;,\label{iner-far-vf}
\label{iner-far-vf}\\
(\delta E)_{rr}\simeq-\frac{\mu^4\sin{(2\omega L+\theta)}}{512\pi^2L}-\frac{\mu^4}{512\pi^3\omega^2 L^3}\label{iner-far-rr}
\label{iner-far-rr}
\eea
with $\theta=\arcsin{\sqrt{1+4\omega^2L^2}}$, indicating that both the vf- and rr-contributions oscillate with the inter-detector separation. The separation-dependence of the total interaction energy is however monotonic:
\bea
\delta E\simeq -\frac{\mu^4}{256\pi^3\omega^2L^3}\;,\label{iner-far}
\label{iner-far}
\eea
as a result of the perfect cancellation of the oscillatory terms in vf- and rr-contributions. Here, one can see that in the far zone,  the vf- and rr-contributions are equally important to the inter-detector interaction, and they jointly result in an attractive force proportional to $L^{-4}$ between the two detectors.

\subsection{Accelerating case}

Now we examine the interaction for two uniformly accelerated detectors. For such an accelerated system, there are generally six typical regions with respect to the three physical parameters, i.e., the inter-detector separation $L$,  the transition wavelength of the detectors $\lambda$, and the acceleration $a$, where analytical results are obtainable. We discover that, in three of the six typical regions, the inter-detector interaction is slightly modified as compared with that in the inertial case, while in others the modifications are very dramatic. These slight and dramatic modifications of the interaction energy can be regarded respectively as a manifestation of weak (local) and strong (long-range) effects of the quantum vacuum altered by the accelerated motion.


\subsubsection{Weak (local) effects of alteration of quantum vacuum}

We find that, in some typical regions, i.e., $L\ll\lambda\ll\lambda_U$, $L\ll\lambda_U\ll\lambda$ and $\lambda\ll L\ll\lambda_U$ with $\lambda_U\equiv2\pi T_U^{-1}$ denoting the thermal wavelength of the Unruh thermal bath, the interaction between two accelerated detectors is slightly modified, as compared with its counterpart in the inertial case. The detailed results are listed  in Tab.~\ref{tab1}.

\begin{center}
\begin{table}[H]
	\renewcommand{\arraystretch}{1.4}
    \caption{Weak (local) effects of alteration in quantum vacuum on the inter-detector interaction energy.}
\begin{threeparttable}
\begin{tabular}{|c| c| c| c|}
 \hline region                                  &  I ($L\ll\lambda\ll\lambda_U$) & II ($L\ll\lambda_U\ll\lambda$) & III ($\lambda\ll L\ll\lambda_U$) \\
  \hline 
  \multirow{2}{*}{$(\delta E)_{vf}$}             &  \multirow{2}{*}{$\frac{\mu^4}{256\pi^3L}-\frac{\mu^4 a^2}{1536\pi^3 \omega^2 L}$}
                                                 &  \multirow{2}{*}{$-\frac{\mu^4}{256\pi^3L}+\frac{7\mu^4a^2L}{6144\pi^3}-\frac{\mu^4\omega^2\ln{\left(\frac{\omega}{a}\right)}}{96\pi a^2 L}$
}
                                                 &  \multirow{2}{*}{$\frac{\mu^4\sin{(2\omega L+\theta)}}{512\pi^2L}-\frac{\mu^4}{512\pi^3\omega^2L^3}-\frac{\mu^4 a^2}{4096\pi^3 \omega^2 L}$} \\
                                                 & & &
                                              \\   
  \multirow{2}{*}{$(\delta E)_{rr}$}             &  $-\frac{\mu^4}{512\pi^2\omega L^2}+\frac{\mu^4}{256\pi^3L}$
                                                 &  \multirow{1}{*}{$-\frac{\mu^4}{512\pi^2\omega L^2}-\frac{\mu^4}{256\pi^3L}$}
                                                 &  \multirow{2}{*}{$-\frac{\mu^4\sin{(2\omega L+\theta)}}{512\pi^2L}-\frac{\mu^4}{512\pi^3\omega^2L^3}-\frac{\mu^4 a^2}{4096\pi^3 \omega^2 L}$}\\
                                                 &  $-\frac{\mu^4 a^2}{1536\pi^3 \omega^2 L}$
                                                 & $+\frac{7\mu^4a^2L}{6144\pi^3}-\frac{\mu^4\omega^2\ln{\left(\frac{\omega}{a}\right)}}{96\pi a^2 L}$
                                                 &  \\ 
  \multirow{2}{*}{vs}                                             &   \multirow{2}{*}{$|\delta E|_{vf}\ll|\delta E|_{rr}$} &  \multirow{2}{*}{$|\delta E|_{vf}\ll|\delta E|_{rr}$}  &  \multirow{2}{*}{$|\delta E|_{vf}\leftrightarrow|\delta E|_{rr}$}\\
                                                 & & &
 \\ 
  \multirow{2}{*}{$\delta E$}                    &  $-\frac{\mu^4}{512\pi^2\omega L^2}+\frac{\mu^4}{128\pi^3L}$
                                                 &  \multirow{1}{*}{$-\frac{\mu^4}{512\pi^2\omega L^2}-\frac{\mu^4}{128\pi^3L}$}
                                                 &  \multirow{2}{*}{$-\frac{\mu^4}{256\pi^3\omega^2L^3}-\frac{\mu^4 a^2}{2048\pi^3 \omega^2 L}$} \\
                                                 &  $-\frac{\mu^4 a^2}{768\pi^3 \omega^2 L}$
                                                 & $+\frac{7\mu^4a^2L}{3072\pi^3}-\frac{\mu^4\omega^2\ln{\left(\frac{\omega}{a}\right)}}{48\pi a^2 L}$
                                                 &  \\
  \hline
\end{tabular}
\begin{tablenotes}
        \footnotesize
        \item[1] Hereafter, $|\delta E|_{vf}\leftrightarrow|\delta E|_{rr}$ denotes that $|\delta E|_{vf}$ can be larger, smaller and even be equal to $|\delta E|_{rr}$, depending on the value of $L$.
\end{tablenotes}
\end{threeparttable}
\label{tab1}
\end{table}
\end{center}

A comparison of the near-region interaction energy, {i.e., the interaction energy} in regions I and II, 
with that of inertial detectors shows that the vf- and rr-contributions, and the total interaction energy are slightly modified by the noninertial motion, and similar to that in the inertial case, the interaction energy dominantly comes from the rr-contribution. Though very slight, there are still important distinctions between the acceleration-induced-modifications in the vf- and rr-contributions in these two regions. In region I where $L\ll\lambda\ll\lambda_U$, the acceleration-induced modification for the vf-contribution appears in the sub-leading order, while that for rr-contribution shows up in the next-sub-leading order. However, in region II where $L\ll \lambda_U\ll\lambda$, the acceleration-independent term which is leading in the  vf-contribution and sub-leading in the rr-contribution remarkably changes sign, suggesting that now the vacuum is essentially altered since the acceleration-independent term is typically characteristic of a vacuum as seen by inertial detectors. As a result, the interaction energy is unexpectedly reinforced,  in the region $L\ll\lambda_U\ll\lambda$,  in the next-leading order which is independent of the acceleration with the acceleration-dependent correction only appearing in the higher order.
More importantly, it is  worth noting that this remarkable  effect of the alteration of the quantum vacuum occurs only when the transition wavelength of the detectors is larger than the characteristic thermal wavelength of the Unruh thermal bath, i.e., $\lambda_U\ll\lambda$.


The comparison of the far-region interaction energy, {i.e., the interaction energy} in region III, 
with its counterpart in the inertial case shows that the accelerated motion brings in equal modifications for both the vf- and rr-contribution, which together slightly strengthen the far-region interaction.

As shown above, the acceleration only induces higher-order modifications to the interaction energy in regions I and III in comparison with the inertial case. It is interesting to note that in regions $L\ll\lambda\ll\lambda_U$, $L\ll\lambda_U\ll\lambda$ or $\lambda\ll L\ll\lambda_U$, the inter-detector separation $L$ is much shorter than the thermal wavelength $\lambda_U$ characteristic of the Unruh thermal bath which an individual detector feels, 
and the two-detector system only probes the local property of the quantum vacuum.\\

\subsubsection{Strong (long-range) effects of alteration of quantum vacuum}

In stark contrast to the small modifications in the short range as is demonstrated above, dramatic modifications to the inter-detector interaction occur when the inter-detector separation $L$ is much larger than the thermal wavelength $\lambda_U$ characteristic of the Unruh thermal bath, meaning that the long-range properties of the quantum vacuum are severely altered by the accelerated motion which then manifests in obvious changes in the  interaction energy. We find that, the changes are especially prominent in regions in which at least one of the typical length scales characterizing the two-detector system, $L$ and $\lambda$, is larger than $\lambda_U$, i.e., $\lambda_U\ll L\ll\lambda$, $\lambda_U\ll\lambda\ll L$, and $\lambda\ll\lambda_U\ll L$, and the corresponding results are listed in Tab.~\ref{tab2}.

\begin{center}
\begin{table}[H]
	\renewcommand{\arraystretch}{1.8}
    \caption{Strong (long-range) effects of alteration in quantum vacuum on the inter-detector interaction.}
\begin{threeparttable}
\setlength{\tabcolsep}{3mm}{
\begin{tabular}{|c|c|c|c|}
   \hline region                                 & IV ($\lambda_U\ll L\ll \lambda$) & V ($\lambda_U\ll \lambda\ll L$) & VI ($\lambda\ll\lambda_U\ll L$) \\
  \hline $(\delta E)_{vf}$                       &  \multicolumn{2}{c|}{$-\frac{\mu^4}{64\pi^3  \omega^2 a L^4}$}
                                                 &  $\frac{\mu^4\ln{(aL)}}{64\pi^2a^3L^4}\sin{(\frac{4\omega\ln{(aL)}}{a}+\psi)}-\frac{\mu^4}{128\pi^3\omega^2aL^4}$\\
         $(\delta E)_{rr}$                       &  \multicolumn{2}{c|}{$-\frac{\mu^4}{64\pi^3  \omega^2 a L^4}$}
                                                 &  $-\frac{\mu^4\ln{(aL)}}{64\pi^2a^3L^4}\sin{(\frac{4\omega\ln{(aL)}}{a}+\psi)}-\frac{\mu^4}{128\pi^3\omega^2aL^4}$\\
         vs                                      &  \multicolumn{2}{c|}{$(\delta E)_{vf}\approx(\delta E)_{rr}$}  &$(\delta E)_{vf}\leftrightarrow(\delta E)_{rr}$ \\
         $\delta E$                              &  \multicolumn{2}{c|}{$-\frac{\mu^4}{32\pi^3\omega^2aL^4}$}
                                                 &  $-\frac{\mu^4}{64\pi^3  \omega^2 a L^4}$\\
  \hline
\end{tabular}}
\begin{tablenotes}
        \footnotesize
        \item[1] $\psi=\arcsin{[1+(4\omega\ln{(aL)}/a)^2]^{-1/2}}$.
      \end{tablenotes}
    \end{threeparttable}
\label{tab2}
\end{table}
\end{center}

In sharp contrast to the results in the inertial case and the weak (local) effects in Tab.~\ref{tab1}, the vf- and rr- contributions and the total interaction energy in regions IV, V and VI are of a completely new behavior, which is surprisingly exclusively acceleration-dependent! In fact, the total interaction energy and the vf- and rr- contributions to it have no acceleration-independent terms at all orders, not just at the leading order as is shown in Tab.~\ref{tab2}. Thus, in these three cases, the vacuum as seen by the accelerated two-detector system with the inter-detector separation much greater than the thermal wavelength of the Unruh thermal bath, i.e., $L\gg\lambda_U$, is drastically different from that as seen by an inertial two-detector system and the accelerated system with the inter-detector separation much smaller than the thermal wavelength of the Unruh thermal bath, i.e., $L\ll\lambda_U$.
In the present case, the acceleration-independent terms in the interaction energy which are typically characteristic of the inertial two-detector system appear to be wiped out completely and the quantum vacuum seems to be in a new phase as perceived by  the accelerated two-detector system. And the long-range parameter signaling this new phase of the quantum vacuum is  $\lambda_U$. These new features are in a sense in accordance with Lee's viewpoint that vacuum {\it ``like any other physical medium, it can carry long-range-order parameters and it may also undergo phase transition."} and his prediction that {\it ``new phenomena, totally unexpected"} may be encountered if the vacuum is indeed altered~\cite{Lee}.


\section{Outlooks and discussions}

As we have demonstrated, the acceleration-induced alteration of a quantum vacuum incarnates itself in remarkable modifications of the inter-detector interaction energy. Taking a ratio between the interaction energy of two detectors in acceleration [referring to the results in Tabs.~\ref{tab1} and \ref{tab2}] and that in inertial motion, $\mathcal{R}_e={{\delta E(a,L)}\/\delta E(0,L)}$, we find that the acceleration-induced alteration of a quantum vacuum generally weakens the near- and far-region inter-detector interaction; however, there exists a characteristic length, $L_0\equiv\sqrt{\lambda\lambda_U}$, which further divides region IV where $\lambda_U\ll L\ll\lambda$ into two subregions, $\lambda_U\ll L\ll L_0$ and $L_0\ll L\ll\lambda$. Remarkably, in $\lambda_U\ll L\ll L_0$ region, $\mathcal{R}_e\approx{8\/\pi^2}({L_0\/L})^2\gg1$, indicating in principle that the near-region interaction energy can be tremendously amplified by the accelerated motion and thus the quantum vacuum can be severely altered. This might also be considered in a sense as an example of vacuum engineering anticipated by Lee~\cite{Lee}. 
{In a real experiment, particles with intrinsic internal energy level structures such as atoms can also be considered as candidates for UDW detectors under discussion here.} Unfortunately, this large amplification is generally difficult to observe 
{with ordinary atoms}, since it happens only when the atoms are accelerated with an extremely high acceleration, in which circumstance, the atoms could be excited and even be ionized.

\begin{figure}[H]
\centering
\includegraphics[width=10cm]{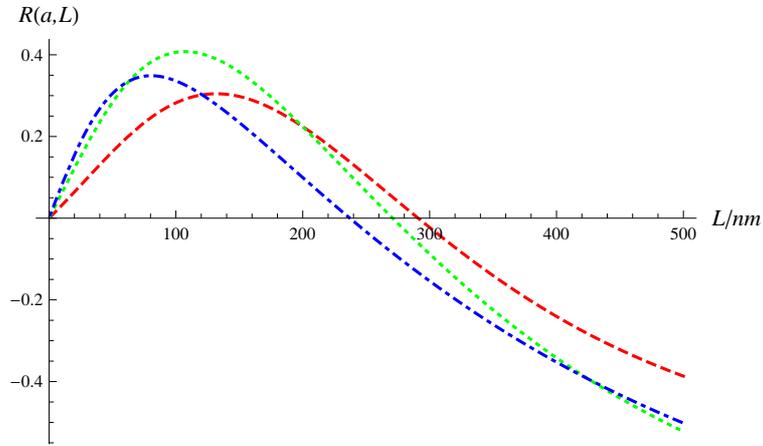}
\caption{\scriptsize Separation-dependence of the interatomic interaction energy of two synchronously uniformly accelerated ground-state sodium atoms. The dashed red, dotted green and dot-dashed blue lines correspond to $a=1.0\times10^{24}m/s^2$, $1.5\times10^{24}m/s^2$, $2.0\times10^{24}m/s^2$ respectively.}
\label{R}
\end{figure}

However, to detect the amplification of the interaction energy in the near region, actually, we do not need to have a very large amplification, and a modification comparable to experimental precision is enough. Here, let us note that there have been several experiments measuring the interaction energy of an atom and surfaces, in which the atom-surface separation ranges from a few to hundreds of nanometers~\cite{Sandoghdar92,Sukenik93,Bender10,Peyrot19} and the best precision is 
{$\sim8\%$}~\cite{Peyrot19}.  In Fig.~\ref{R},  we plot, the results of our numerical computation of the relative modification of the interaction energy for two uniformly accelerated sodium atoms as compared to the inertial case, $R(a,L)=\frac{\delta E(a,L)}{\delta E(0,L)}-1$, as a function of  the interatomic separation in such a range, i.e.,  $L\in[1,500]nm$. This figure shows that, the relative modification induced by accelerated motion can be either positive or negative when the interatomic separation is relatively short or large, indicating that the interaction energy can be either amplified or weakened, and when $a\sim1.5\times10^{24}m/s^2$ and $L\sim110nm$, the relative modification can be as large as $\sim40\%$. Remarkably, when the two atoms are 
{$\sim255nm$} apart and accelerated with 
{$a\sim3.0\times10^{23}m/s^2$}, the acceleration-induced-amplification has already reached 
{$ 8\%$}.  This acceleration is of three orders lower than that required for the detection of the Unruh effect~\cite{Crispino08,Chen99}.

However, so far, a direct measurement of the interaction energy of two atoms hundreds of nanometers apart still seems to be a tough task due to the fact that the interaction of two atoms with such a separation is much smaller than that of an atom and a surface. But with two Rydberg atoms (atoms with large principal quantum number) which exhibit very strong interactions, the direct measurement of the near-region interaction energy of two atoms a few microns apart has been realized~\cite{Beguin13}. It has been reported that, by working in the regime where the Rabi frequency of a single $^{87}Rb$ atom for excitation to the Rydberg state is comparable to the near-region interaction, partial Rydberg blockade is observed to take place, and as a result the near-region interaction energy of the two atoms can be extracted by comparing quantitatively the time-dependent populations of the various two-atom states which exhibit coherent oscillations with several frequencies with a simple model based on the optical Bloch equations~\cite{Beguin13}.
An experimentally detectable interaction energy of atoms with a larger separation means a lower  acceleration required  for the detection of the modification of it.
For two Rydberg atoms separated by $L\in[4,20]\mu m$, a distance in which the near-region interaction energy is experimentally detectable~\cite{Beguin13}, we find that the modification  of the interaction energy induced by an acceleration, $a\sim2\times 10^{21}m/s^2$, which is $10^5$ times smaller than that required for the detection of the Unruh effect~\cite{Crispino08,Chen99}, reaches the current experimental precision~\cite{Beguin13}. This acceleration is still much larger than $\sim10^{15}m/s^2$, the largest acceleration of neutral atoms, which, to the best of our knowledge, can be currently realized with strong short-pulse laser fields~\cite{Echimann09}. Thus the hope for a future experimental detection of the effects we reported in this paper that reveal long-range properties of the quantum vacuum lies in the progress of technology for accelerating neutral atoms and the improvement of precision of experimental measurement of the interatomic interaction energy.

\begin{acknowledgments}

We would like to thank Jiawei Hu for helpful discussions.
This work was supported in part by the NSFC under Grants No. 11690034, No. 12075084, No. 11875172, No. 12047551 and {No. 12105061}, and the K. C. Wong Magna Fund in Ningbo University.

\end{acknowledgments}

\begin{appendix}
{\section{The derivations of the vf-contribution and rr-contribution to the inter-detector interaction energy, Eqs.~(\ref{vf}) and (\ref{rr}). }
\label{derivation}
With the fourth-order DDC formalism~\cite{Zhou21}, the contribution of vacuum fluctuations to the interaction energy of two detectors in their ground states 
{and in synchronized motion} is given by,
\bea\label{2vf}
(\delta E)_{vf}
&=&2i\mu^4\int_{\tau_0}^{\tau}d\tau_1\int_{\tau_0}^{\tau_1}d\tau_2\int_{\tau_0}^{\tau_2}d\tau_3C^F(x_A(\tau),x_B(\tau_3))\chi^F(x_A(\tau_1),x_B(\tau_2))\chi^A(\tau,\tau_1)\chi^B(\tau_2,\tau_3)\nonumber\\
&&+\text{$A\Leftrightarrow B$ {\it term}}\;,
\eea
and that of the radiation reaction of the detectors by
\bea\label{2rr}
(\delta E)_{rr}
&=&2i\mu^4\int_{\tau_0}^{\tau}d\tau_1\int_{\tau_0}^{\tau_1}d\tau_2\int_{\tau_0}^{\tau_2}d\tau_3\chi^F(x_A(\tau),x_B(\tau_3))\chi^F(x_A(\tau_1),x_B(\tau_2))C^A(\tau,\tau_1)\chi^B(\tau_2,\tau_3)\nonumber\\
&&+2i\mu^4\int_{\tau_0}^{\tau}d\tau_1\int_{\tau_0}^{\tau_1}d\tau_2\int_{\tau_0}^{\tau_2}d\tau_3\chi^F(x_A(\tau_1),x_B(\tau_3))\chi^F(x_B(\tau_2),x_A(\tau))C^A(\tau,\tau_1)\chi^B(\tau_2,\tau_3)\nonumber\\
&&+2i\mu^4\int_{\tau_0}^{\tau}d\tau_1\int_{\tau_0}^{\tau_1}d\tau_2\int_{\tau_0}^{\tau_2}d\tau_3\chi^F(x_A(\tau_3),x_B(\tau_2))\chi^F(x_B(\tau_1),x_A(\tau))C^A(\tau,\tau_3)\chi^B(\tau_1,\tau_2)\nonumber\\
&&+2i\mu^4\int_{\tau_0}^{\tau}d\tau_1\int_{\tau_0}^{\tau}d\tau_2\int_{\tau_0}^{\tau_2}d\tau_3\chi^F(x_A(\tau_2),x_B(\tau_3))\chi^F(x_A(\tau),x_B(\tau_1))\chi^A(\tau,\tau_2)C^B(\tau_1,\tau_3)\nonumber\\
&&+2i\mu^4\int_{\tau_0}^{\tau}d\tau_1\int_{\tau_0}^{\tau_1}d\tau_2\int_{\tau_0}^{\tau}d\tau_3C^F(x_B(\tau_2),x_A(\tau_3))\chi^F(x_B(\tau_1),x_A(\tau))\chi^A(\tau_3,\tau)\chi^B(\tau_1,\tau_2)\nonumber\\
&&+2i\mu^4\int_{\tau_0}^{\tau}d\tau_1\int_{\tau_0}^{\tau_1}d\tau_2\int_{\tau_0}^{\tau_1}d\tau_3\chi^F(x_A(\tau),x_B(\tau_3))\chi^F(x_A(\tau_1),x_B(\tau_2))C^A(\tau,\tau_1)\chi^B(\tau_3,\tau_2)\nonumber\\
&&+\text{$A\Leftrightarrow B$ {\it terms}}\;,
\eea
where $C^{\xi}$ and $\chi^{\xi}$ are respectively the symmetric and antisymmetric statistical functions of the detectors in their ground states, which are defined as
\bea
C^{\xi}(\tau,\tau')&\equiv&\frac{1}{2}\langle g_{\xi}|\{R^{\xi,f}_{2}(\tau),R^{\xi,f}_{2}(\tau')\}|g_{\xi}\rangle,\label{Cd}\\
\chi^{\xi}(\tau,\tau')&\equiv&\frac{1}{2}\langle g_{\xi}|[R^{\xi,f}_{2}(\tau),R^{\xi,f}_{2}(\tau')]|g_{\xi}\rangle,\label{Chid}
\eea
with 
\bea\label{R2f}
R^{\xi,f}_{2}(\tau)=\frac{i}{2}\left[R^{\xi}_{-}(\tau_0)e^{-i\omega_{\xi}(\tau-\tau_0)}-R^{\xi}_{+}(\tau_0)e^{i\omega_{\xi}(\tau-\tau_0)}\right]
\eea
being the free part of the atomic operator $R^{\xi}_{2}(\tau)$, and $C^F$ and $\chi^F$ are respectively the symmetric correlation function and the linear susceptibility of the scalar fields in the vacuum state $|0\rangle$, which are defined as
\bea
C^F(x_A(\tau),x_B(\tau'))&\equiv&\frac{1}{2}\langle 0|\{\phi^{f}(x_A(\tau)),\phi^{f}(x_B(\tau'))\}|0\rangle,\label{fieldC}\\
\chi^F(x_A(\tau),x_B(\tau'))&\equiv&\frac{1}{2}\langle 0|[\phi^{f}(x_A(\tau)),\phi^{f}(x_B(\tau'))]|0\rangle,\label{fieldChi}
\eea
with
\bea
\phi^f(x)=\int d^3{\mathbf{k}}\  g_{\mathbf{k}}\left[a_{\mathbf{k}}(t_0)e^{-i\omega_{\mathbf{k}}(t-t_0)}e^{i\mathbf{k}\cdot\mathbf{x}}
+a^{\dag}_{\mathbf{k}}(t_0)e^{i\omega_{\mathbf{k}}(t-t_0)}e^{-i\mathbf{k}\cdot \mathbf{x}}\right]\label{free-field}
\eea
being the free scalar field operator.}

{For the two detectors moving along trajectories Eq.~(\ref{trajectories}), the symmetric correlation function and the linear susceptibility of the scalar fields defined in Eqs. (\ref{fieldC}) and (\ref{fieldChi}) can be further simplified to
\bea
C^F(x_A(\tau),x_B(\tau'))&=&\frac{1}{2\pi}\int_{-\infty}^{\infty}d\omega'\  \mathcal{G}(\omega',a,L)\cos{[\omega'(\tau-\tau')]}\;,\label{cf}\\
\chi^F(x_A(\tau),x_B(\tau'))&=&-\frac{i}{2\pi}\int_{-\infty}^{\infty}d\omega'\  \mathcal{G}(\omega',a,L)\sin{[\omega'(\tau-\tau')]}\label{chif}
\eea
where 
\bea\label{cfchif2}
\mathcal{G}(\omega',a,L)=\frac{\sin{\left[\frac{2\omega'}{a}\sinh^{-1}\left(\frac{aL}{2}\right)\right]}}{2\pi L\sqrt{1+a^2L^2/4}}\frac{1}{1-e^{-2\pi\omega'/a}}\;.
\eea
For two-level detectors, the two statistical functions $C^{\xi}$ and $\chi^{\xi}$ defined in Eqs.~(\ref{Cd}) and (\ref{Chid}) can be further expressed as
\bea
C^{\xi}(\tau,\tau')&=&\frac{1}{4}\cos{[\omega_{\xi}(\tau-\tau')]},\label{Cdtwolevel}\\
\chi^{\xi}(\tau,\tau')&=&-\frac{i}{4}\sin{[\omega_{\xi}(\tau-\tau')]}.\label{Cdtwolevel}
\eea
Use Eqs.~(\ref{cf})-(\ref{Cdtwolevel}) in Eqs.~(\ref{2vf}) and (\ref{2rr}), perform the triple integrations on $\tau_3$, $\tau_2$ and $\tau_1$, and then we obtain the vf- and rr-contribution to the inter-detector interaction energy, Eqs.~(\ref{vf}) and (\ref{rr}).}

\end{appendix}



\end{spacing}

\end{document}